\documentclass[
pra,
aps,
twocolumn,
superscriptaddress,
longbibliography,
floatfix,
notitlepage]{revtex4-2}

\usepackage[utf8x]{inputenc}
\usepackage[english]{babel}
\usepackage[T1]{fontenc}
\usepackage{lmodern}

\usepackage{amsmath,amsfonts,amssymb,amsthm,bm,times,dcolumn,dsfont}
\usepackage{mathrsfs}
\usepackage{microtype}
\usepackage{braket}
\usepackage{gensymb} 

\usepackage{xcolor}
\usepackage[colorlinks={true}, citecolor={blue}, filecolor={blue}, linkcolor={blue}, urlcolor={blue}]{hyperref}
\usepackage{graphicx,color}
\usepackage[caption=false]{subfig}
\usepackage{float}

\usepackage[]{algpseudocode}
\usepackage{algorithm}

\begin{document}

\title{Super-Resolution Imaging with Multiparameter Quantum Metrology in Passive Remote Sensing}

\author{Emre Köse}
  \email{saban-emre.koese@uni-tuebingen.de }
\affiliation{Institut f\"{u}r Theoretische Physik, Eberhard Karls Universit\"{a}t T\"{u}bingen, 72076 T\"{u}bingen, Germany}

\author{Daniel Braun}
  \email{daniel.braun@uni-tuebingen.de}
\affiliation{Institut f\"{u}r Theoretische Physik, Eberhard Karls Universit\"{a}t T\"{u}bingen, 72076 T\"{u}bingen, Germany}

    \date{\today}
\begin{abstract}
We study super-resolution imaging theoretically using a distant $n$-mode
interferometer in the  microwave regime for passive remote sensing,
used e.g., for satellites like the "soil moisture and ocean salinity (SMOS)"
mission to observe the surface of the Earth. We give a complete quantum mechanical analysis of multiparameter estimation of the temperatures on the
source plane. We find the optimal detection modes by 
combining incoming modes with an optimized unitary that enables  the most informative measurement based on photon
counting in the detection modes and saturates the quantum Cramér-Rao bound from the symmetric logarithmic derivative for the parameter set of
temperatures. In our numerical analysis, we achieved a
quantum-enhanced super-resolution by reconstructing an image using the
maximum likelihood estimator with a pixel size of 3 km, which is ten
times smaller than the spatial resolution of SMOS with comparable
parameters. Further, we find the optimized unitary for uniform
temperature distribution on the source plane, with the
temperatures corresponding to the average temperatures of the
image. Even though the corresponding unitary was not optimized for the
specific image, it still gives a super-resolution compared to local
measurement scenarios for the theoretically possible maximum number of
measurements.  
\end{abstract}
\maketitle
\section{Introduction}
The technology of imaging is currently undergoing a rapid evolution
both due to enhanced computational techniques
\cite{bhandari_computational_2022}, and due to
insights from quantum information processing and quantum metrology. It
has become clear that the paradigmatic resolution limit found by Abbe
and Rayleigh based on the interference of classical waves, set  by the
wavelength of the light, is not the ultimate fundamental bound if the
quantum nature of light is taken into account.  In quantum optics it
was realized already in the 1960s in the context of the explication of
the Hanbury-Brown Twiss effect
\cite{hanbury_brown_test_1956,fano_quantum_1961} that fundamentally 
the interference of light should be considered in Hilbert space and
can lead to higher order correlations that contain information beyond
the first order correlations relevant for the interference patterns
of classical electromagnetic waves. Experimentally,
super-resolution was demonstrated by Hell in 1994 
\cite{hell_breaking_1994,hell_farfield_2007}, who resolved a
molecule with nanometer resolution with light
in the optical domain by decoration of the molecule with point-like
emitters and quenching them selectively.
Theoretical work used early on the techniques of optimal parameter
estimation to estimate the ultimate sensitivities of radar and in fact
led to the development of quantum parameter estimation theory
\cite{helstrom_detection_1967,helstrom_quantum_1969,helstrom_cramerrao_1973,helstrom_estimation_1970}.
Much later, quantum parameter estimation theory was applied to
determine optimal detection modes and ultimate sensitivities for
arbitrary parameters encoded in the quantum state of Gaussian light
\cite{pinel_ultimate_2012,pinel_quantum_2013}.  In 2016, Tsang and
coworkers wrote a seminal paper that considered the problem of
ultimate resolution as quantum parameter estimation problem for the
distance between the two sources. They found that the Quantum Fisher
information (QFI) that sets the ultimate bound remains finite for two
point sources of low, identical intensity in the limit of vanishing
separation, whereas the classical Fisher information linked to
intensity measurements in direct imaging vanishes.  A large amount of
theoretical
\cite{tsang_quantum_2019,zhou_modern_2019,sorelli_momentbased_2021,rehacek_multiparameter_2017,napoli_superresolution_2019,nair_farfield_2016,lupo_ultimate_2016,larson_resurgence_2018,kurdzialek_superresolution_2021,kolobov_quantum_2000,ang_quantum_2017,tsang_quantum_2011,bisketzi_quantum_2019,bojer_quantitative_2021,datta_subrayleigh_2020,dealmeida_discrimination_2021,liang_coherence_2021,tsang_subdiffraction_2017,tsang_quantum_2015,karuseichyk_resolving_2022,lupo_quantum_2020,bojer_quantitative_2021,gottesman_longerbaseline_2012,khabiboulline_optical_2019,wang_superresolution_2021}
and experimental research
\cite{backlund_fundamental_2018,mazelanik_optical_2021,paur_achieving_2016,pushkina_superresolution_2021,boucher_spatial_2020,sorelli_optimal_2021}
followed that corroborated and generalized this insight. 

Most of these works concentrated on estimating one or few parameters,
however, typically linked to geometrical information like the spatial separation or position of point sources and,  {in some cases, optical phase imaging, i.e., the joint estimation of the phases with respect to a reference mode \cite{humphreys_quantum_2013,gagatsos_gaussian_2016,knott_local_2016,pezze_optimal_2017}.}
While this led to important insights and solid evidence that in many
situations quantum parameter estimation techniques can enhance
resolution beyond the classical diffraction limit, imaging typically
aims not at recovering information about the separation, or, more
generally, the spatial position, of point sources.  Rather, in a
typical image, the scene is covered by pixels of known locations and
one wants to know for each pixel the intensity of the
source in that point, its spectral composition, polarization etc.
Since an image consists typically of many pixels, imaging
is then inherently a (quantum-) many-parameter estimation problem, and
corresponding techniques should be applied to obtain the best
possible quality of an image re-construction based on the gathered
measurement results. 

In this work we go an important step in this direction in passive
remote sensing of Earth in the micro-wave domain, building 
on our previous work \cite{kose_quantumenhanced_2022}.   
{Here the state of
the art is interferometric antenna synthesis, with which a large
effective antenna can be formed from a set of small antennas, with
corresponding enhanced 
resolution. For example, the SMOS (``Soil Moisture and Ocean
Salinity'') satellite is an interferometer with a Y-shaped array of 
69 antenna with each arm has a length of around 4 m
\cite{anterrieu_resolving_2004,corbella_visibility_2004,levine_synthetic_1999,thompson_interferometry_2017}.
It achieves a resolution of 
about $d\simeq 35$\,km, from a distance $R\simeq 758$\,km  above the
surface of Earth, by measuring the thermal noise in a narrow frequency
band of electromagnetic fields (1420-1427 MHz, central wavelength
$\lambda\simeq 21$\,cm). The electric fields are sampled in real-time,
filtered and interfered numerically, implementing thus 
purely classical interference. The diffraction limit analogous to the
ones by Abbe and Rayleigh is given here by  
the van Cittert-Zernike theorem 
\cite{vancittert_wahrscheinliche_1934,zernike_concept_1938,braun_generalization_2016},
$d\simeq\lambda R/\Delta x_{ij}$, where $\Delta x_{ij}$ is the maximal
spatial separation between two antennas. From the interferometric data
one can, via inverse spatial Fourier Transform, estimate the local
brightness temperatures $T_\text{eff}$ on the surface of Earth with
resolution $d$, and from these, with appropriate models, the soil
moisture and ocean salinity.  This information is of great importance
for the geosciences, monitoring of Earth, climate modelling, flood
predictions, and many more.  Driven by these applications, there is the
desire to enhance the spatial resolution, but simply increasing the
size of the satellite becomes unpractical, and lowering its orbit
reduces its life-time.  \\ 

Here we show that with appropriate techniques from multiparameter
quantum estimation theory, one can reconstruct images of Earth with
roughly a factor of 10 times better spatial resolution than SMOS with a
satellite of comparable size. We demonstrate this with images of
up to 30 pixels, for which we show that they can be reconstructed faithfully with a pixel size of 3\, km. Instead of local measurement of the incoming modes of the interferometer, we combine the modes with a unitary transformation that
enables  non-local measurements. We find the optimal unitary matrix that minimizes the scalar classical Cramér Rao bound \cite{albarelli_perspective_2020} for the classical Fisher information matrix for the chosen measurements 
contracted with a weight matrix.
The corresponding unitary matrix can be decomposed into phase shifters and at most $n(n-1)/2$ beam splitters, as is well-known from linear optical quantum computing  \cite{kok_linear_2007}.  This allows us to quantum-program optimal measurement schemes for imaging.  Note that contrary to classical computational imaging \cite{bhandari_computational_2022} the quantum computation for this new kind of ``quantum-computational imaging'' is done before the measurements.  

Multiparameter quantum estimation theory is by itself a rapidly evolving field. Recently, there have been many different works, 
e.g., multiparameter estimation of 
several phases \cite{humphreys_quantum_2013}, estimation of all three components of a magnetic field 
\cite{baumgratz_quantum_2016}, optimal estimation  of the Bloch vector components 
of a qubit \cite{bagan_optimal_2006}, multiparameter estimation from Markovian dynamics \cite{guta_information_2017}, etc. (see the review article \cite{szczykulska_multiparameter_2016}). For a limited sample size, like in passive sensing, it is crucial to simultaneously estimate the image's parameters. The multi-parameter quantum Cram\'er-Rao bound can in general not be saturated.  Optimal measurement linked to different parameters do typically not commute and hence lead to incompatible measurements. Once the commutation on average is satisfied, the quantum limit is asymptotically attainable \cite{ragy_compatibility_2016}.

We build on our previous work
\cite{braun_generalization_2016,braun_fouriercorrelation_2018,kose_quantumenhanced_2022},
where we showed that thermal fluctuations of the microscopic 
currents lead to Gaussian states of the microwave field and hence
allow one to use the QCRB for Gaussian states
\cite{liu_quantum_2020,pinel_ultimate_2012,pinel_quantum_2013,shapiro_quantum_2009}, 
As before we assume that only the current densities at the surface of
Earth contribute and neglect the cosmic microwave background as well
as dditional technical noises \cite{oh_quantum_2021,gessner_superresolution_2020,len_resolution_2020}.

{We organize the rest of the article as follows. In Section
  \ref{theory}, we 
  introduce the quantum state received by the $n$-mode interferometer,
  as well as the quantum Fisher information (QFI), the symmetric
  logarithmic derivative (SLD, and the corresponding quantum Cramer
  Rao lower bound (QCRB). Further, we present the optimal POVM (positive-operator-valued
 measure), which minimizes the most informative bound for the
 multiparameter estimation. In Section \ref{result}, first, we discuss
 the simple problem as a benchmark considering two-pixel sources with
 the two-mode interferometer. We analyze the quantum advantage with
 the optimal unitary compared to local measurement scenarios. Second,
 we increase the number of pixels by considering a 1D array of sources
 with a 1D array interferometer. We examine how 
 closely we can approache the quantum limit 
 of sensitivity 
 with our parameter set. Third, we consider a 2D source image with a
 2D array interferometer. Using the maximum likelihood estimator, we
 reconstruct the image for the POVMs with the optimized unitary
 specific to the image, the optimized unitary for uniform temperature
 distribution, and local measurements. 
 We conclude in Section \ref{conclude}.}

\section{Theory}
\label{theory}
\subsection{The State Received by $n$-mode Interferometer} 
In previous work \cite{kose_quantumenhanced_2022}, we analyzed the quantum state radiated from current current distribution $\mathbf{j}(\mathbf{r},t)$ \cite{ braun_fouriercorrelation_2018,blow_continuum_1990,mandel_optical_1996,glauber_coherent_1963,scully_quantum_1999,loudon_quantum_1974,kubo_fluctuationdissipation_1966,savasta_light_2002,sharkov_passive_2011,landau_statistical_1980,carminati_nearfield_1999} on the source plane. We show that the state of the incoming modes of the $n$-mode interferometer from these radiated sources can be modeled as circularly symmetric Gaussian states with a partial coherence, which encodes the information of position and amplitudes distribution of the sources. Then after the scattering process \cite{zmuidzinas_cramer_2003,zmuidzinas_thermal_2003} from the interferometer the partially coherent state received in the $n$ modes is represented by  
\begin{equation}
    \rho=\int \text{d}^{2n}\beta \Phi(\{\beta_i\})|\{\beta_i\}\rangle\langle\{\beta_i\}|,
\end{equation}
where $|\{\beta_i\}\rangle$ is a multi-mode coherent state for spatial antenna modes, $\{\beta_i\} = \beta_1, \beta_2, ... \beta_n$, and
\begin{equation}
    \Phi(\{\beta_i\})=\frac{1}{\pi^{n} \operatorname{det} \Gamma} e^{-\bar\beta^\dagger \Gamma ^{-1}\bar \beta}.
    \end{equation}
with $\bar \beta ^T = (\beta_1, \beta_2...\beta_n )$ is the Sudarshan-Glauber representation, and $\text{d}^{2n}\beta\equiv \text{d}\Re{\beta_1}\text{d}\Im{\beta_1}\ldots \text{d}\Re{\beta_n}\text{d}\Im{\beta_n}$. The matrix $\Gamma$ is the coherence matrix for $n$ antenna modes and its elements are defined as $\Gamma_{ij} = \braket{\hat b^\dagger_i \hat b_j}$. Considering the sources of these fields are generated by random current distribution on the source plane and assuming that each antenna has the same polarization direction $\hat e_l$ and they filter incoming fields with same frequency $\omega_0$ with a bandwidth $B$, then one finds a relation between $\braket{b^\dagger_i b_j}$ and the average current density distribution on the source plane as \cite{kose_quantumenhanced_2022}
\begin{equation}
	\begin{split}
	\braket{\hat b^\dagger_i \hat b_j} &=K\int d^{3} {r}\;\frac{\braket{|\tilde{{j}}_{t,l}\left(\mathbf{r},{\omega}\right)|^2}e^{i\omega_0(|\mathbf{r}-\mathbf{r}_j|-|\mathbf{r}-\mathbf{r}_i|)/c}}{|\mathbf{r}-\mathbf{r}_i||\mathbf{r}-\mathbf{r}_j|}\\&\times\mathrm{sinc}\left[\frac{B}{2c}(|\mathbf{r}-\mathbf{r}_j|-|\mathbf{r}-\mathbf{r}_i|)\right],
	\end{split}
	\label{eq3}
\end{equation}
where $d^3r$ is the integral over the source volume, $\mathbf{r}_i$ is the location of the detector for received modes in the detection plane and $\text{sinc}(x) = \sin(x)/x$. 
$\tilde{{j}}_{t,l}\left(\mathbf{r},{\omega}\right)$ is the Fourier transform of the locally transverse component of the current density ${\mathbf{j}}\left(\mathbf{r},t\right)$ and '$l$' stands for the component parallel to the source plane. Considering $R$ as the distance between source and detection planes, we can parametrize the integral over Earth's surface as $\mathbf{r}= (x,y,R)$ with respect to the coordinate system of the detection plane. Assuming that we are in the far field regime $|\Delta\mathbf{r}_{ij}|\ll R$, where $\Delta\mathbf{r}_{ij}= \mathbf{r}_j-\mathbf{r}_i$ is the distance between two antennas, we approximate $|\mathbf{r}-\mathbf{r}_j|-|\mathbf{r}-\mathbf{r}_i| \approx \Delta\mathbf{r}_{ij}\cdot \mathbf{r}/|\mathbf{r|}$. In the denominator, we approximate $|\mathbf{r}-\mathbf{r}_i| \approx R/\cos \tilde\theta(x,y) $ with $\tilde\theta(x,y)$ the polar angle between the $z$-axis and the vector $(x,y,R)$. We find the relation of the average amplitude of current density distribution to brightness temperature as $T_{\mathrm{B}}(x,y)$ by $\braket{|\tilde{{j}}_{t,i}\left(\mathbf{r},{\omega}\right)|^2} = K_1 T_{\mathrm{B}}(x,y)\cos\tilde \theta(x,y)\delta(z-R)$, where $K_1 = 32 \tau_c k_B /(3 l_c^3 \mu _0 c)$. Further, one can define the effective temperature as $T_{\mathrm{eff}}(x,y)\equiv T_{\mathrm{B}}(x,y) \cos^3\tilde \theta(x,y)$. We include an extra constant prefactor $\mu$ for the additional losses, which can be justified by tracing out modes of losses "$\hat c$" into which photons might scatter by writing $\hat b = \sqrt{\mu} \hat{\tilde b} + \sqrt{1-\mu} \hat c$. 
Compared to the actual physical temperature, the brillance temperature is additionally modified by the albedo of the surface from which important information such as the water content of the surface or the salinity of ocean water can be extracted. For simplicity we simply work with the physical temperatures in the following, i.e., set $T_B(x,y)=T(x,y)$. 
Following these assumptions and dropping the $\sim$ from $\hat{\tilde b} $, we simplify Eq. (\ref{eq3}) as 
\begin{equation}
	\begin{split}
	\braket{\hat b^\dagger_i \hat b_j}&=\frac{\mu\kappa}{R^2}\int dxdy\;T_{\mathrm{eff}}(x,y)e^{2\pi i\left(v_x^{ij}x +v_y^{ij}y \right)},
	\end{split}
	\label{eqxi}
\end{equation}
We introduced a new constant $
\kappa =  K_1K \equiv {2k_B}/{(\pi \hbar \omega_0)} $ where $\kappa$
has the dimension of inverse temperature with SI-units
"$(1/\mathrm{K})$" and 
$v_y^{ij} = {\Delta x_{ij} }/({\lambda R}),\quad v_x^{ij} ={\Delta y_{ij} }/({\lambda R})$ with $\omega_0/c = 2\pi /\lambda$. Considering the parameters of SMOS we find $\kappa = 9.4 $ 1/K. The SMOS has a Y shape where each arm has a length of almost 4 m. Therefore, it is reasonable to set maximum baselines $\Delta x_{\max} = \Delta y_{\max} $ around 10 m.

\begin{figure}[t!]
	\centering
	\includegraphics[width=0.99\linewidth]{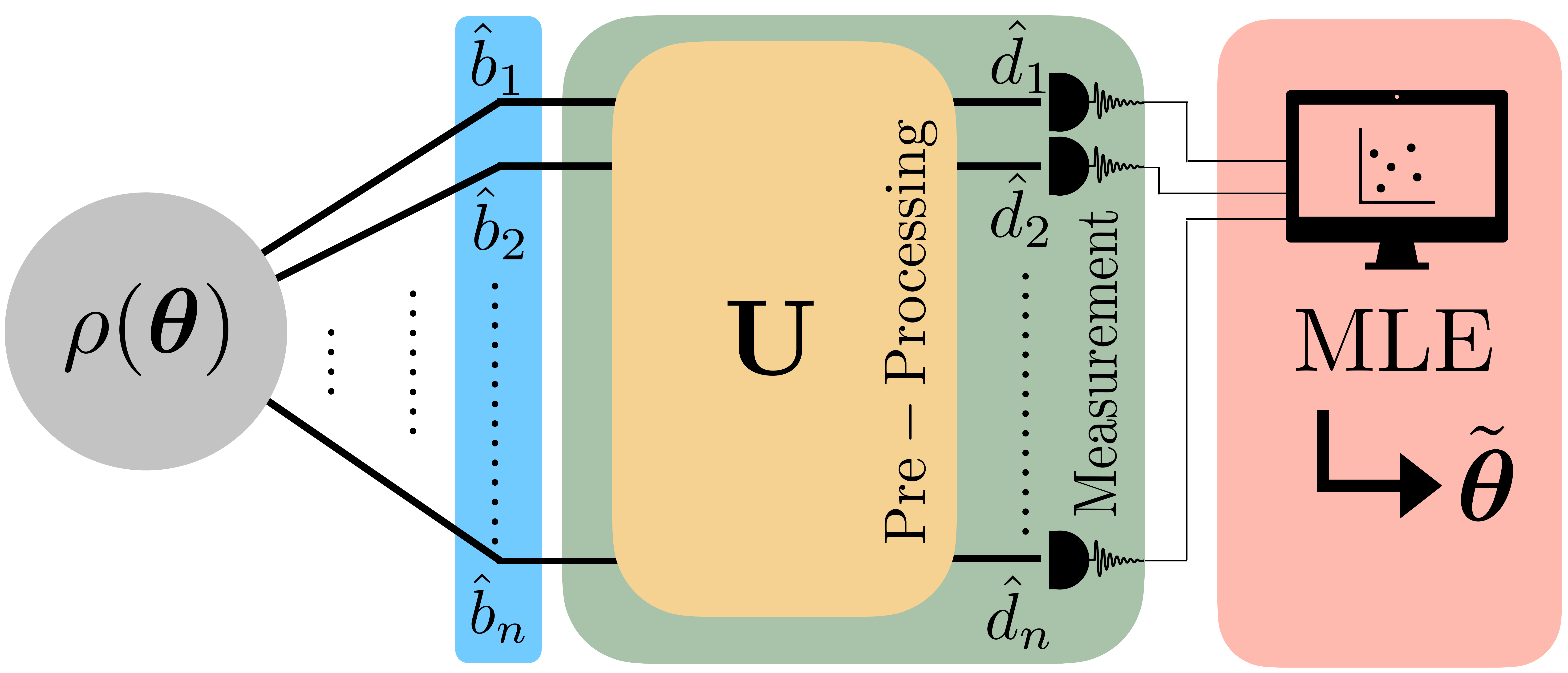}
	\caption{The Gaussian state $\rho(\boldsymbol{\theta})$  of the $n$-mode interferometer contains spatial and radiometric information from current density sources. The incoming modes $\hat b_i$ are combined with an optimized $\mathbf{U}$ to have detection modes $\hat d_i$ of the photon counting measurement. For experimental realization, one can decompose $\mathbf{U}$ into $SU(2)$ group elements similar to optical quantum computing, i.e., using beam splitters and phase shifters. After the measurements, one estimates the parameter set using an estimator function such as a maximum likelihood estimator (MLE).}
	\label{interferometer}
\end{figure}

\subsection{Estimation Theory of the Sources}
\label{sec2c}
\textit{Quantum Cram\'er-Rao Bound:} For a quantum state $\rho_{\boldsymbol{\theta}}$ that depends on a vector of $l$ parameters $ \boldsymbol{\theta} = (\theta_1,\theta_2,...,\theta_l)^T$, an ultimate lower bound of an unbiased estimator of the parameter set is given by the quantum Cram\'er-Rao (CR) bound, which states that the co-variance matrix of any such estimator is equal or greater than the inverse of the QFI matrix (in the sense that their difference is a positive-semidefinite matrix). The classical Cram\'er-Rao bound (CCRB) from measurement is lower bounded by the quantum Cram\'er-Rao bound (QCRB) ~\cite{helstrom_quantum_1969,helstrom_detection_1967,szczykulska_multiparameter_2016} given by  
\begin{equation}
	\operatorname{Cov}(\tilde{\boldsymbol{\theta}}) \geqslant \mathscr{F}(\boldsymbol{\theta})^{-1}, \quad \mathscr{F}_{{i j}}(\boldsymbol{\theta})=\frac{1}{2} \operatorname{Tr}\left(\rho_{\boldsymbol{\theta}}\left\{\mathscr{L}_{i}, \mathscr{L}_{j}\right\}\right),\label{multi}
\end{equation}
where $\operatorname{Cov}(\tilde{\boldsymbol{\theta}}) $ is a covariance matrix for the locally unbiased estimator $ \tilde{\boldsymbol{\theta}}$ \cite{ragy_compatibility_2016,sidhu_geometric_2020}, the $\{\cdot,\cdot \}$ means the anti-commutator, and $\mathscr{L}_{i}$ is the symmetric logarithmic derivative (SLD) related to parameter $i$, which is defined similarly to the single-parameter case, $\frac{1}{2}\left(\mathscr{L}_{i} \rho_{\boldsymbol{\theta}}+\rho_{\boldsymbol{\theta}} \mathscr{L}_{i}\right)=\partial_{{i}} \rho_{\boldsymbol{\theta}} .$ The SLD and the elements of the QFI matrix are given in Ref.~\cite{gao_bounds_2014} for any Gaussian state. The SLD can be written as
 \begin{equation}
 \mathscr{L}_{i}=\frac{1}{2} \mathfrak{M}_{\alpha \beta, \gamma \delta}^{-1}\left(\partial_{i} \Sigma^{\gamma \delta}\right)\left(\mathbf{b}_{\alpha} \mathbf{b}_{\beta}-\Sigma^{\alpha \beta}\right),
 \end{equation}
 where the summation convention is used, and in our case, that the mean displacement of the Gaussian state is zero. Covariance matrix elements are $\Sigma_{i j}=\frac{1}{2} \operatorname{Tr}\left[\rho\left({\mathbf{b}}_i {\mathbf{b}}_j+{\mathbf{b}}_j {\mathbf{b}}_i\right)\right]$, with $\mathbf{b}=\left[b_1, b_1^{\dagger}, b_2, b_2^{\dagger}, \ldots b_n, b_n^{\dagger}\right]$ ~\cite{braun_precision_2014,adesso_continuous_2014,gao_bounds_2014,olivares_quantum_2012,pinel_ultimate_2012,weedbrook_gaussian_2012}. Then the elements of the QFI matrix in \cite{gao_bounds_2014} become
\begin{equation}
\mathscr{F}_{i j}=\frac{1}{2} \mathfrak{M}_{\alpha \beta, \gamma \delta}^{-1} \partial_{j} \Sigma^{\alpha \beta} \partial_{i} \Sigma^{\gamma \delta},
\end{equation}
where $\mathfrak{M} \equiv \Sigma \otimes \Sigma+\frac{1}{4}\Omega \otimes \Omega$,
and $\Omega=\bigoplus_{k=1}^{n} i \sigma_{y}$.
Using the properties of the Gaussian state (circularly symmetric and with zero mean) we can write the SLD for $n$ mode interferometers as \cite{kose_quantumenhanced_2022}
\begin{equation}
    \begin{split}
      \mathscr{L}_{i} = \sum_j^n g^{j}_i \hat{b}^\dagger _j \hat{b}_j + \sum _{j<k}^n (g^{jk}_i\hat{b}^\dagger _j \hat{b}_k +  (g^{jk}_{i})^*\hat{b}^\dagger _k \hat{b}_j) +\mathrm{C},
    \end{split}
	\label{eq:Sld}
\end{equation}
where C is a constant term that can be dropped for diagonalization purposes. In the single parameter case, the optimal POVM is the set of projectors onto eigenstates of $\mathscr{L}_i$. It allows one to saturate the QCRB in the limit of infinitely many measurements using maximum likelihood estimation \cite{helstrom_detection_1967,braunstein_statistical_1994,paris_quantum_2009}. To find the POVMs from the SLD, we construct a Hermitian matrix $\mathbf{M}_i$ whose diagonal elements are real-valued functions which are defined as $g^{j}_i \equiv \mathfrak{M}_{\alpha \beta, \gamma \delta}^{-1}\left(\partial_{i} \Sigma^{\gamma \delta}\right) $ with $\alpha = 2j$ and $\beta = 2j-1$. The off-diagonal elements are complex-valued functions and defined as $g^{jk}_i =\mathfrak{M}_{\alpha \beta, \gamma \delta}^{-1}\left(\partial_{i} \Sigma^{\gamma \delta}\right) $ with $\alpha = 2j$ and $\beta = 2k-1$ and $k>j$. By introducing a new set for the field operators such that, $\bar{\mathbf{b}}^\dagger \equiv {\left[\hat b_1^\dagger, \hat b_2^\dagger, ..., \hat b_n^\dagger\right]}$ and $\bar{\mathbf{b}} \equiv {\left[ \hat b_1, \hat b_2, ..., \hat b_n\right]^T}$, we write the SLD in the following form
\begin{equation}
    \begin{split}
      \mathscr{L}_{i} = \bar{\mathbf{b}}^\dagger \mathbf{M}_i \bar{\mathbf{b}}.
    \end{split}
\end{equation}
As $\mathbf{M}_i$ is a Hermitian matrix it can be unitarily
diagonalized by $\mathbf{M}_i =\mathbf{V}^\dagger_i \mathbf{D}_i
\mathbf{V}_i $ with $\mathbf{V}_i^\dagger \mathbf{V}_i = \mathbb{I}$. A new set of operators can be defined as $\bar{\mathbf{d}}_i^\dagger =
\bar{\mathbf{b}}^\dagger \mathbf{V}_i^\dagger$ where
$\bar{\mathbf{d}}_i^\dagger = {\left[\hat d_{i1}^\dagger, \hat
 d_{i2}^\dagger, ..., \hat d_{in}^\dagger\right]}$. The optimal
POVM for the single parameter case ($i=1$, which we drop in the
following) can be found as a set of projectors in the Fock basis $\{\ket{m_1,m_2,...,m_n}\bra{m_1,m_2,...,m_n}\}_{\{m_1,m_2...m_n\}}$ of the $\hat d_l $ with $\hat d^\dagger_l \hat d_l\ket{m_1,m_2,...,m_n} = m_l\ket{m_1,m_2,...,m_n}$, where $l\in \{1,...,n\}$. The $\hat{d}_l$ will be called "detection modes." {By introducing a positive weight matrix $\boldsymbol{w}$, one can define the scalar inequalities from the matrix valued QCRB as $\operatorname{Tr}(\boldsymbol{w}\operatorname{Cov}(\tilde{\boldsymbol{\theta}})) \geqslant \operatorname{Tr}(\boldsymbol{w}\mathscr{F}(\boldsymbol{\theta})^{-1}) \equiv C^S(\boldsymbol{\theta},\boldsymbol{w})$. Contrary to the single parameter case, the multiparameter QCRB can generally not be saturated. Holevo realized this problem and proposed a tighter and more fundamental bound \cite{holevo_statistical_1973} $C^H(\boldsymbol{\theta},\boldsymbol{w})$, which is upper bounded by $2C^S(\boldsymbol{\theta},\boldsymbol{w})$ \cite{tsang_quantum_2020,albarelli_perspective_2020}. If the SLD operators for different parameters commute on average $\operatorname{Tr}(\rho_{\boldsymbol{\theta}}[L_i, L_j])=0$, then the Holevo-CRB is equivalent to the QCRB, 
and  the QCRB for multiparameter estimation can be saturated asymptotically with a collective measurement in the limit of an infinitely large number of copies $\rho_{\boldsymbol{\theta}}^{\otimes{N}}$ \cite{ragy_compatibility_2016,albarelli_perspective_2020}.} The standard deviation of the estimator decreases proportionally to $1/\sqrt{{N}}$ for the sample size of ${N}$. The SMOS satellite travels at a constant speed of around $v\simeq 7$ km/s. It takes time $\tau = L/v$ to fly at a distance $L$. Each sample has a lower bound for the detection time given by $t_D\simeq 1/B $. In practice, the practical detection time might be much larger due to, e.g.~deadtimes of the sensors, slow electronics, etc. In addition, zero temperature of the detector and modes $\hat b_i$ is implicitly assumed in our calculations but would require cooling down to temperatures much smaller than $\hbar \omega_0$. If the actual detection time is $t_D^\text{eff}$, the maximum sample size becomes ${N}=\tau/t_D^\text{eff}$.

\textit{Most Informative Bound for Multiparameter Metrology:} The most informative bound minimizes the classical scalar Cramer Rao bound over all the possible POVMs. In the single parameter case, from the diagonalization of the SLD, we see that one needs to combine the incoming modes with a unitary transformation to saturate the QCRB single parameter case. This transformation, even for a single parameter, depends on the parameter itself. In the multiparameter case, any of these specific unitary transformations for a specific parameter usually gives a more significant mean square error for the remaining parameters. Using the clue from the SLD structure, we drop the index "$i$" from the unitary transformation of the modes and minimize the scalar bound of the classical Fisher information matrix for multiparameter estimation over all possible unitaries. Then, a new set of operators for the detection modes can be defined as $\bar{\mathbf{d}} =\mathbf{U}
\bar{\mathbf{b}} $ where
$\bar{\mathbf{d}}^T = {\left[\hat d_{1}, \hat
d_{2}, ..., \hat d_{n}\right]}$, where $\mathbf{U}$ is the corresponding unitary transformation of the field modes. The average values of the elements of the new coherence matrix $\tilde \Gamma$ can be found by using $\hat d_i = \sum_l U_{il} \hat b_l $ as 
\begin{equation}
   \tilde \Gamma_{ij}=  \langle d_i^{\dagger} d_j \rangle = \sum_{kl} U^*_{ik}U_{jl}\langle b_k^{\dagger} b_l \rangle.
\end{equation}
Then we will have the probabilities after measurement $P(m_1,..m_n| \theta_1,\theta_2,...,\theta_l) $ as
\begin{equation}
    \begin{split}
        P(\{m_k\}| \boldsymbol{\theta})&=\int \text{d}^{2n} {\delta} \tilde \Phi(\{\delta_i\}) |\langle \{m_k\}|\{\delta_i\}\rangle|^2 ,\\&=\int \text{d}^{2n} {\delta} \tilde \Phi(\{\delta_i\})  \prod_i e^{-|\delta_i|^2}\frac{|\delta_i|^{2m_i}}{m_i!}.
    \end{split}
\end{equation}
where $|\{\delta_i\}\rangle $ is a coherent state of the detection modes and $ \tilde \Phi(\{\delta_i\})$ is the Sudarshan-Glauber function for the state of the detection modes. Due to the linear transformation from $\bar{ \mathbf{b}}$ to $\bar {\mathbf{d}}$ it is still a Gaussian. It is difficult to evaluate the integral of $P(\{m_k\}| \boldsymbol{\theta})$ for all possible values of $m_k$ and keep track of all possible combinations of photon number counts, both numerically and experimentally. Hence, instead of considering projections on the complete Fock basis as POVMs, we choose the POVMs with at most one photon per measurement and limit ourselves to 
$\sum_k m_k\leq 1 $. 
Clearly, the resulting information loss is negligible for light that from the very beginning is very faint, with at most one photon per mode, but can be important for stronger light sources, for which one should try to resolve the photon numbers. For thermal microwave sources at room temperature, we have of the order of 10 photons per mode.  We see below that even without resolving their number we can already largely surpass the classical resolution limit, but there is room for further improvement by going beyond the single-photon detection scheme that we analyse in the following.\\ 

The selected POVM elements  of single photon detection are 
\begin{equation}
    \begin{split}
        &\Pi_0 = \ket{0,0,...,0}\bra{0,0,...,0}, \\
        &\Pi_k = \ket{0,0,...,1_k,...,0}\bra{0,0,...,1_k,...,0}, \\
        &\Pi_{n+1} = \mathbb{I}-\sum_{l=0}^{n}\Pi_l,
    \end{split}
\end{equation}
where the last element ($n+1$) ensures  $\sum_{l=0}^{n+1} \Pi_l = \mathbb{I}$. The measurement probability of no photon in any interferometer mode becomes
\begin{equation}
    \begin{split}
        P_0(\boldsymbol{\theta}) &=\frac{1}{\pi^{n} \operatorname{det} \tilde \Gamma}\int \text{d}^{2n} {\delta} e^{-\boldsymbol{\delta}^\dagger (\tilde \Gamma^{-1} +\mathbb{I})\boldsymbol{\delta}}\\ &= \frac{1}{\operatorname{det} (\tilde \Gamma +\mathbb{I})}.
    \end{split}
	\label{eq:13}
\end{equation}
The single photon detection probabilities in each mode of the interferometer follow as 
\begin{equation}
    \begin{split}
        P_k(\boldsymbol{\theta}) &=\frac{1}{\pi^{n} \operatorname{det} \tilde \Gamma}\int \text{d}^{2n} {\delta} e^{-\boldsymbol{\delta}^\dagger (\tilde \Gamma^{-1} +\mathbb{I})\boldsymbol{\delta}}|\delta_k|^2\\ &= \frac{[(\tilde \Gamma^{-1} +\mathbb{I})^{-1}]_{kk}}{\operatorname{det} (\tilde \Gamma +\mathbb{I})}.
    \end{split}
	\label{eq:14}
\end{equation}
The probability to find more than a single photon per measurement, can be found as 
\begin{equation}
    P_{n+1}(\boldsymbol{\theta}) = \mathbb{I}-\sum_{k=0}^{n}P_k.
\end{equation}
We also show the first derivative of the probability distributions of no photon detection from measurements analytically to be given by
\begin{equation}
    \begin{split}
        \frac{\partial P_0(\boldsymbol{\theta})}{\partial \theta_i} &= \left(\frac{1}{\operatorname{det} (\tilde \Gamma +\mathbb{I})}\right) \operatorname{Tr}\left(-(\tilde \Gamma +\mathbb{I})^{-1}\frac{\partial \tilde \Gamma }{\partial  \theta_i}\right).
    \end{split}
	\label{eq:15}
\end{equation}
The first derivative for at most single photon detection for all modes becomes
\begin{equation}
    \begin{split}
        \frac{\partial P_k(\boldsymbol{\theta})}{\partial \theta_i} &= \left(\frac{1}{\operatorname{det} (\tilde \Gamma +\mathbb{I})}\right)\\&\times\left[{[(\tilde \Gamma^{-1} +\mathbb{I})^{-1}\tilde \Gamma^{-1}\frac{\partial \tilde \Gamma }{\partial  \theta_i}\tilde \Gamma^{-1}(\tilde \Gamma^{-1} +\mathbb{I})^{-1}]_{kk}}\right. \\&- \left.  [(\tilde \Gamma^{-1} +\mathbb{I})^{-1}]_{kk} \operatorname{Tr}\left((\tilde \Gamma +\mathbb{I})^{-1}\frac{\partial \tilde \Gamma }{\partial  \theta_i}\right)\right] .
    \end{split}
	\label{eq:17}
\end{equation}
Finally, using all Eqs. (\ref{eq:13}-\ref{eq:17}), the elements of the classical Fisher information can be found from  
\begin{equation}
	\mathcal{F}_{ij} = \sum_l^{n+1} \frac{1}{P_l(\boldsymbol{\theta})}  \frac{\partial P_l(\boldsymbol{\theta})}{\partial \theta_i}   \frac{\partial P_l(\boldsymbol{\theta})}{\partial \theta_j} .
	\label{cFI}
\end{equation}
The most informative bound \cite{albarelli_perspective_2020} in this case is the bound minimized over all possible unitary matrices 
\begin{equation}
	\operatorname{Tr}\left[\boldsymbol{w}\operatorname{Cov}(\tilde{\boldsymbol{\theta}})\right] \geq \min _{\mathbf {U }}\left[\operatorname{Tr}\left[\boldsymbol{w} \mathcal{F}^{-1}(\boldsymbol{\theta})\right]\right].
	\label{MIB}
\end{equation}
For simplicity, we will consider $\boldsymbol{w} = \mathbb{I}$.

\textit{Maximum Likelihood Estimation:} Maximum likelihood estimators are widely used in estimation theory and play an essential role in interpreting the Cramér-Rao theorem \cite{myung_tutorial_2003,paris_quantum_2004}. One can estimate the set of parameters with a given probability distribution with some observed data. The likelihood function is given by  
$l(\boldsymbol{\theta}) = \prod_k ^{n+1} (P_k(\boldsymbol{\theta}))^{N_k} $,
where the total number of samples is given by $N = \sum_k^{n+1} N_k$ with $N_k$ realizations of outcome $k$. Since the logarithm is a monotonously increasing function, the log of the likelihood function is maximized by the same parameter vector $\boldsymbol{\theta}$. Thus, the maximum likelihood estimator (MLE) $\boldsymbol{\hat\theta}_{\mathrm{mle}}$ is a value  of $\boldsymbol{\theta}$ that maximizes the log-likelihood $\mathcal{L}(\boldsymbol{\theta})= \log (l(\boldsymbol{\theta}))$,
\begin{equation}
	\boldsymbol{\hat{\theta}}_{\text {mle }}=\underset{\boldsymbol{\theta} \in \Theta}{\arg \max } \mathcal{L}(\boldsymbol{\theta}),
	\end{equation}
where the max is taken over the entire parameter space $\Theta$. For sufficiently large sample size, $N\rightarrow \infty$, $\boldsymbol{\hat{\theta}}_{\text {mle }}$ converges to the true value of the parameter set $\boldsymbol{\theta}$.

\section{Results: Estimation of source temperatures}
\label{result}

We partition the source of the electromagnetic field on the surface of Earth into square pixels of size $a$ and effective pixel temperature $T_i$, located under the interferometer in the $x,y$ plane at distance $R$ from the satellite. 
We are interested in estimating the temperature distribution
\begin{equation}
    T_{\mathrm{eff}}(x,y)= \sum_i T_i \mathrm{Box}(x-x_i,y-y_i),
\end{equation}
where $\mathrm{Box}(x,y)$ is defined as 
\begin{equation}
    \operatorname{Box}(x, y) \triangleq \begin{cases}1 & |x|\leq \frac{a}{2}\quad \mathrm{and} \quad|y|\leq \frac{a}{2}\\ 0 & \mathrm{else}\end{cases}.
\end{equation}
We assume that all the other parameters are known to sufficiently large precision. The diagonal elements of the coherence matrix ($\Gamma$) of Gaussian states becomes 
\begin{equation}
    \langle \hat b_k^\dagger  \hat b_k \rangle = \frac{\mu\kappa a^2 }{R^2} \sum_i^p T_i,
\end{equation}
and the off-diagonal elements are
\begin{equation}
\begin{split}
	\langle \hat b_k^\dagger  \hat b _l \rangle  =&  \frac{\mu\kappa a^2 \eta_{kl} }{R^{2}} \sum_i^p T_i e^{2 \pi i\left(v^{x}_{kl} x_i+v^{y}_{kl} y_i\right)},
\end{split}
\end{equation}
where $k\neq l$ and we defined $\eta_{kl} \equiv\operatorname{sinc}(v^{x}_{kl}a)\operatorname{sinc}(v^{y}_{kl}a) $. The number of pixels along the $\hat x $ and $\hat y$ axis is $p_x$ and $p_y$, respectively, and the number of detection modes along these axes $n_x$ and $n_y$, respectively. In total, we have $p = p_xp_y$ pixels on the surface and $n = n_xn_y$ detectors in the detection plane of which each measures one detection mode. We set the number of detection modes equal to the number of pixels in the source plane, $n=p$, to leave no redundant parameter for the estimation, and 
  use $n_x=p_x$ and $n_y=p_y$.

\subsection{Resolution of two pixel sources}

Let us start with two pixels (pixel-1 and pixel-2) with temperatures $T_1$ and $T_2$ in the source plane with pixel size $a$. Our goal is to estimate the temperatures of each source. We set the central locations of these two sources in the source plane to $(-a/2,0,R)$ and $(a/2,0,R)$, i.e.~both are on an axis parallel to the $\hat x$-axis without any distance between them. In the detection plane, we have two detection modes $\hat d_1$ and $\hat d_2$ with detectors  centered 
at positions $(-\Delta x/2,0,0)$ and $(\Delta x /2,0,0)$ on the $\hat x$-axis, respectively. In our previous work \cite{kose_quantumenhanced_2022}, we showed that if the mean photon numbers in each received mode of the two-mode interferometer, with circular symmetric Gaussian state, are identical ($\langle b_1^\dagger b_1\rangle = \langle b_2^\dagger b_2\rangle$), then the SLDs for $T_1$ and $T_2$ commute on average $\operatorname{Tr}(\rho_{\boldsymbol{\theta}}[L_i, L_j])=0$. Thus the QCRB and Holevo-CRB are equivalent $C^S(\boldsymbol{\theta},\boldsymbol{w})\equiv C^H(\boldsymbol{\theta},\boldsymbol{w})$. For each parameter, the matrix $\mathbf{M}_i$ from the SLD with $i\in \{T_1,T_2\}$, is of the form 
\begin{equation}
	\mathbf{M}_i=\left[\begin{array}{cccc}
	g_{1}^i & |g_{2}^{i}| e^{i\phi_i}  \\
	|g_{2}^{i}| e^{-i\phi_i} & g_{1}^i 
	\end{array}\right],
\end{equation}
where the $\phi_i$, in general, depend on both $T_1$ and $T_2$. The $\phi_{1}$ and $\phi_{2}$ differ
for single parameter estimation of $T_1 $ and $T_2$. The unitary that diagonalizes each SLD is found as 
\begin{equation}
	\mathbf{U}_i=\frac{1}{\sqrt{2}}\left[\begin{array}{cc}
	1 & e^{i \phi_i} \\
	1 & -e^{i \phi_i}
	\end{array}\right].
\end{equation}
Since the unitary is parametrized with a single parameter, we can drop the index $i$ and find the $\phi$ that gives the most informative bound for joint estimation of both $T_1$ and $T_2$. 
\begin{figure}[t!]
	\centering
	\includegraphics[width=1.0\linewidth]{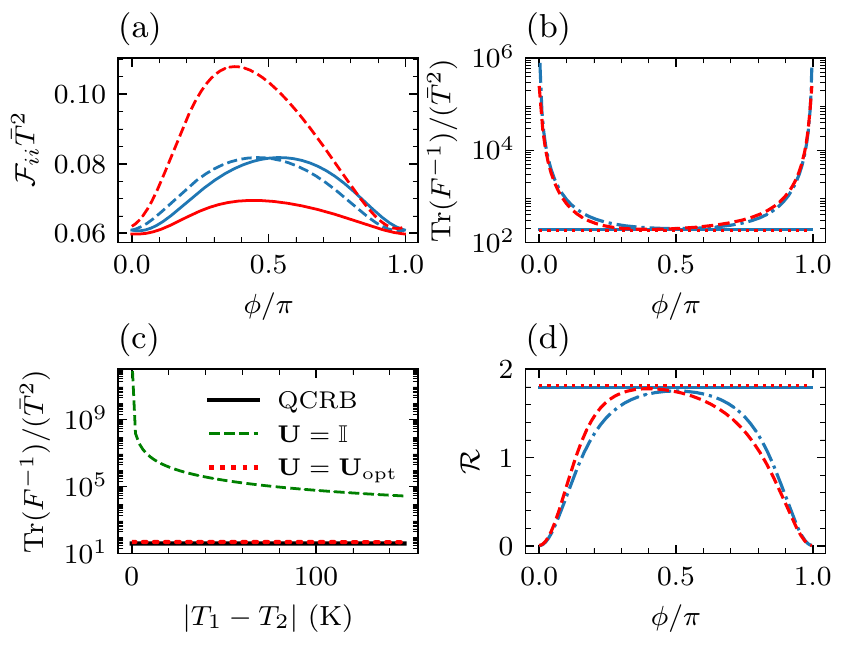}
	\caption{Temperature estimation of two pixels.  (a) The diagonal elements of the Fisher information matrix (dimensionless) as a function of $\phi$. The dashed curves are for $\mathcal{F}_{T_1}$, solid curves are for $\mathcal{F}_{T_2}$. (b) The scalar CRBs as a function of $\phi$ scaled with the average temperature  $\bar T$ square of the sources. The flat lines are for the QCRBs, the others are the CCRBs for the measurement. (c) The scalar CRBs as a function of temperature difference of two pixels. Solid black is for QCRB, red dotted is for CCRB for optimized $\phi$, and green dashed curve is for the scalar CCRB for local measurement considering $\mathbf{U} = \mathbb{I}$. (d) The 
	gain factor of the estimate $\mathcal{R}$ as function of $\phi$. The
	flat lines are from the QCRBs, the others are the CCRBs from the measurement.
	In figures (a), (b), and (d), the blue curves are for uniform temperature, $T_1 = T_2 = 300$ K and red curves are for non-uniform temperatures, $T_1 = 400$ K and $T_2 = 200$ K. The source size is $a=4$ km. The average temperature in all figures is $\bar T=300$ K and $\mu = 0.5$.}
	\label{figtwopixel}
\end{figure}
In Fig.~\ref{figtwopixel}(a), we plot the diagonal elements of the CFI matrix in Eq. (\ref{cFI}) as a function of $\phi$. If $T_1$ (dashed) and $T_2$ (solid) are equal, $T_1 =T_2$ (blue lines), 
a diagonal element   $\mathcal{F}_1 $ or $\mathcal{F}_2$, can be obtained by mirroring the other with respect to $\phi=\pi/2$. 
For different temperatures, $T_1 > T_2$ (red lines), the CFI matrix elements are not symmetric anymore. We observe that $\max(\mathcal{F}_1) > \max(\mathcal{F}_2)$, and their difference is {related} to temperature changes, means that we can estimate the pixel with higher temperature better. We keep the average temperature ($\bar T$) constant in both cases. In both cases we have the maximum value of CFI matrix elements $\max(\mathcal{F}_1) =\max(\mathcal{F}_2)$ at different $\phi$ and diagonalizes the SLD for each parameter for single parameter estimation. 

\begin{figure*}[t]
	\centering
	\begin{tabular}{ccc}
		\qquad\includegraphics[width=5.2cm]{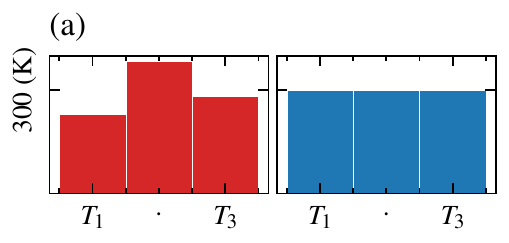}& 
		\includegraphics[width=5.2cm]{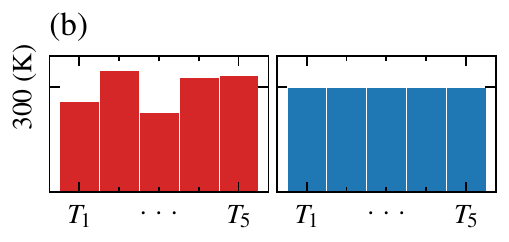}&
		\includegraphics[width=5.2cm]{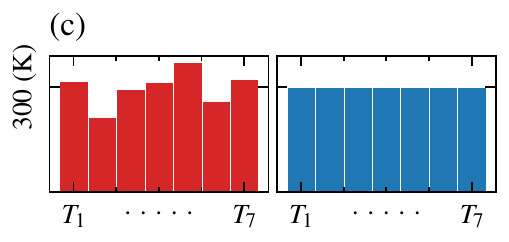}\\
\includegraphics[width=6.0cm]{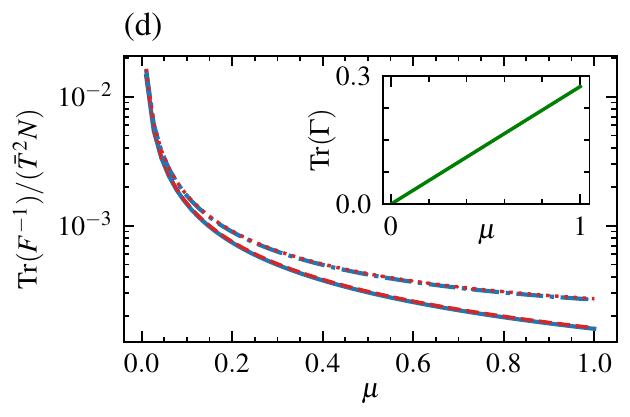} & 
\includegraphics[width=5.5cm]{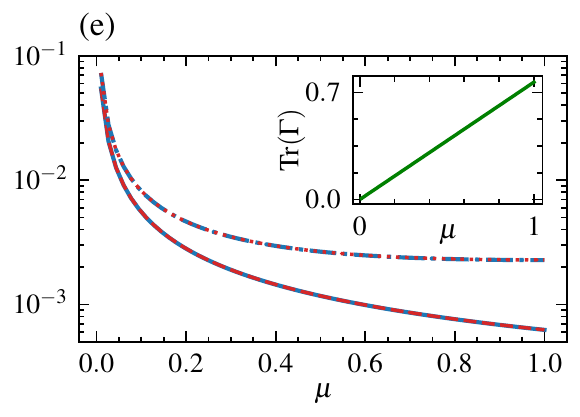} &
\includegraphics[width=5.5cm]{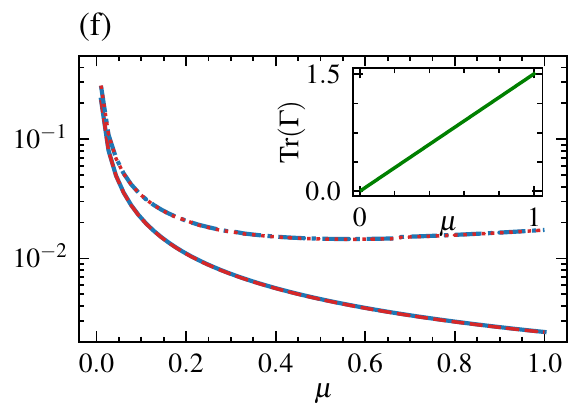}\\ 
\includegraphics[width=6.0cm]{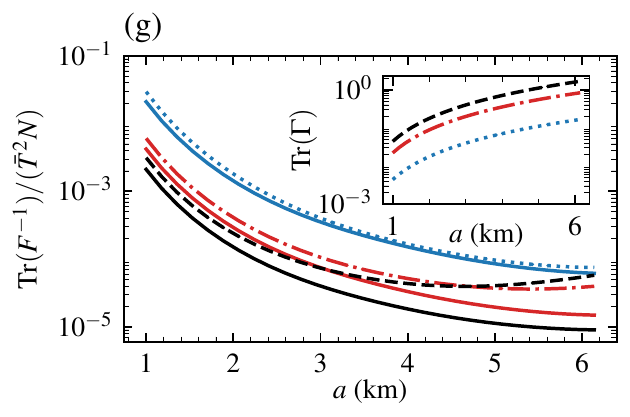}&
\includegraphics[width=5.5cm]{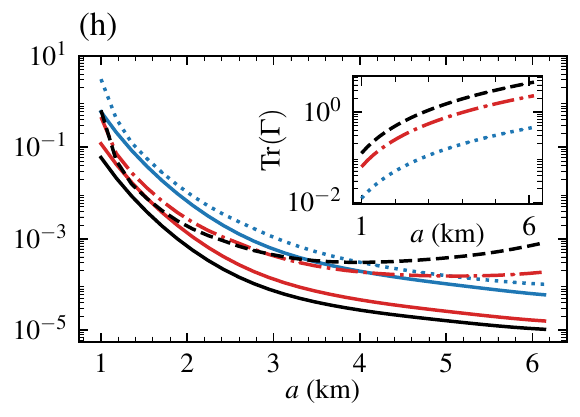}& 
\includegraphics[width=5.5cm]{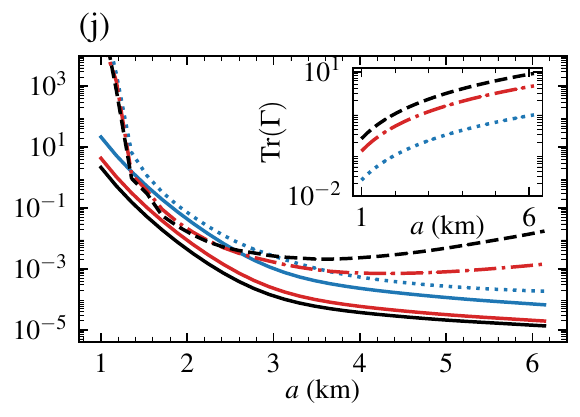} 
	\end{tabular} 
	\caption{(a-c) The temperature distribution of the 1D pixels with uniform temperature
          (blue bars) and different temperatures 
          (red bars) for $3,5$, and $7$ pixels of the sources from (a) to (c), respectively. The scalar CRBs (dimensionless) as a function of $\mu$ (d-f) for the number of source pixels corresponding to (a-c). The solid blue and dashed red curves describe the QCRBs, and dotted red curves and dash dotted blue curves CCRBs for uniform and random temperature configurations. The insets show the total photon number as a function of $\mu$ with a solid green line. The pixel size for (d) to (f) is 2.5 km. 
          The figures from (g) to (j) show the scalar CRBs as a function of the source size $a$. The blue, red, and black lines correspond to different $\mu = (0.1,0.5,1.0)$, respectively. The solid lines represent the QCRBs, and dashed, dash-dotted and dotted lines the scalar CCRBs of single photon measurements with optimized unitary specific to different pixel configurations. The insets show the total photon number in the detector as a function of pixel size $a$, with the corresponding color of different $\mu$. The average temperatures are assumed to be $\bar T = 300$ K, and the sample size is set to be $N=10^6$.}
	\label{fig1Dpixel}
\end{figure*}

In Fig \ref{figtwopixel}(b), we plot $ \mathrm{Tr}({F}^{-1})/(\bar T^2)$ as a function of $\phi$ for $T_1=T_2$ (blue) and $T_1>T_2$ (red) temperature configurations. The scalar QCRBs are given by solid blue ($T_1=T_2$) and dotted red ($T_1>T_2$) 
flat lines, respectively. We see that for $T_1 = T_2$ (dot-dashed blue curve), we have the minimum of the scalar CCRB at $\phi = 0.5 \pi$, and for $T_1>T_2$ (dashed red curve), the minimum value is slightly shifted to the left. In both cases, the QCRBs are saturated. We see that the magnitude of scalar QCRBs for $T_1= T_2$ and $T_1 > T_2$ are close to each other if we keep the same $\bar T$ in both configurations. We also observe that $ \mathrm{Tr}(\mathcal{F}^{-1})/(\bar T^2)$ for $T_1>T_2$ (dashed red curve) at $\phi = \pi/2$ is still close to the QCRB (red dotted 
flat line). Even though to saturate the QCRB, $\phi$ must depend on the temperatures of all pixels, one can find the $\phi$ for $T_1 = T_2 = \bar T$ and use it to estimate different temperature configurations ($T_1 > T_2$). 

In Fig.~\ref{figtwopixel}(c), we compare the most informative bound for optimal $\phi$ with the CCRB of local measurement (i.e. $\mathbf{U}=\mathbb{I}$) for joint estimation of $T_1$ and $T_2$ for a single measurement. We see that the dimensionless CCRB for the local measurement (green dashed line) goes to "$\infty$" when the two sources have the same temperature. 
For a temperature difference around $\sim 10$ K, it is around $\sim 10^6$, which is almost $\sim 10^4$ times larger than 
for a optimal non-local measurement using $\mathbf{U}_{\mathrm{opt}}$ (red dotted line). We also see that the optimal unitary saturates the QCRBs (solid black line). The bounds given in Fig.~\ref{figtwopixel} are for a single measurement ($N=1$) and reduce by a factor $N$ for $N$ independent measurements. 

One can wonder what is the advantage of joint estimation of parameters over single parameter estimation? To answer that question, we can define the gain factor of the joint estimate \cite{nichols_multiparameter_2018,yousefjani_estimating_2017},
\begin{equation}
	\mathcal{R}=p \sum_i^{p} \frac{1 / F_{i i}}{\operatorname{Tr}\left(F^{-1}\right)},
\end{equation}
where $p$ is the total number of the parameters we want to estimate. The $F$ stands for both the QFI matrix $\mathscr{F}$ and the CFI matrix $\mathcal{F}$. The gain factor $\mathcal{R}$ is upper bounded by $p$ ($0<\mathcal{R} \leq p$), 
where the factor $p$ arises from the fact that for $p$ single parameter estimations the number of samples available for each parameter is reduced by a factor $p$ compared to the total sample size, as different optimal measurements are typically required for different parameters. 
Since we have only two parameters to estimate ($T_1$ and $T_2$), the upper bound of the gain factor becomes $\mathcal{R} \leq 2$. If the gain factor is smaller than one, $\mathcal{R}<1$, then we do not have any advantage from joint estimation. In Fig \ref{figtwopixel}(d), we show the gain factor $\mathcal{R}$ of the estimation as a function of $\phi$. It is close to $2$ for the scalar QCRBs of $T_1 = T_2$ (solid blue) and $T_1 > T_2$ (dotted red straight lines). Furthermore, this advantage is achieved by the optimized unitary for CCRBs of $T_1= T_2$ (dot-dashed blue curve) and $T_1>T_2$ (dashed red curve), and we have almost twice the advantage compared to single parameter estimation.
\\\\\\

\subsection{Resolution of 1D array of pixel sources}
We next consider a 1D array of pixels aligned parallel to the detector modes on the $\hat x$ axis ($p_x=n_x$ and $p_y = n_y = 1$). The size $a$ of a pixel is the same for all pixels, and the separation between the two nearest pixels vanishes. The central position of each pixel is given by $\tilde x_j = (2j-p_x-1)a/2$, and the position of detector $k$ is $x_k =(2k-n_x-1) \Delta x_{\max}/n_x$, where $j \in \{1,...,p_x\}$ and  $k \in \{1,...,n_x\}$. The parameters that we want to estimate are the temperatures of each pixel given by a vector $\bm{\theta} = \{ T_1, T_2,...T_{p_x}\}$. 

The unitary $\mathbf{U}$ becomes a $n_x\times n_x$ matrix, and we need $n_x^2$ real parameters. Varying 
independently
all the parameters of $\mathbf{U}$ to find a minimum for our cost function is a difficult task. Therefore,
for $n>2$, we 
use 
a steepest decent algorithm to minimize the most informative bound in Eq. (\ref{MIB}). An efficient algorithm to minimize a given cost function with an argument of the Lie group of unitary matrices $U(n)$ is proposed in Ref \cite{abrudan_conjugate_2009}. The unitary group $U(n)$ is a real Lie group of dimension $n^2$. In each iteration step, the conjugate gradient (CG) algorithm moves towards a minimum along the geodesic on the Riemannian manifold, corresponding to a straight line in Euclidean space. We explain the details of the CG algorithm adapted from Refs. \cite{abrudan_conjugate_2009,abrudan_efficient_2007,abrudan_steepest_2008,abrudan_efficient_2008} in Appendix \ref{appen}. These types of algorithms are widely used in classical communication systems. In this paper, we use the algorithm to optimize the POVM to achieve the quantum limit for imaging in passive remote sensing. We verified numerically that for our choice of the parameter set, the SLDs for different parameters commute on average over the corresponding quantum state for the $n$-mode interferometer.

In Fig.~\ref{fig1Dpixel}, we analyze the QCRB and the CCRB for different numbers of source pixels $p_{x}$ (3, 5, and 7). The average temperatures are fixed to  $\bar T= 300$ K for both random temperature distributions (left-red bars) and the uniform temperature distribution of the pixel sources (right-blue bars). From Figs \ref{fig1Dpixel}(d) to \ref{fig1Dpixel}(f), we show how the classical bounds from our measurement with optimized unitary change as a function of $\mu$, insets show the changes of the corresponding total photon numbers as a function of $\mu$ in each configuration. Since the total mean photon number of the detection modes (solid green lines) decreases with $\mu$ and tends to $\mathrm{Tr}(\Gamma)\ll 1$, the POVMs of single photon detections (red dotted and blue dash-dotted) saturate the QCRBs (red dashed and solid blue) for different and uniform temperature configurations, respectively. When $\mathrm{Tr}(\Gamma) $ gets close to one, we see that the gap between the QCRB and the CCRB for single photon measurement with optimized unitary ($\mathbf{U}_\mathrm{opt}$) increases. Additionally, the QCRBs decrease as the number of photons increases with $\mu$, which means more photons from each pixel increase the QFI of the parameters. Thus, one needs to perform photon-number measurements rather than just single-photon to achieve the QCRB in this limit. Increasing the number of pixels $p$ increases the total photon number on the interferometer. Thus the gap between the QCRBs and the CCRBs for measurement with optimized ($\mathbf{U}_\mathrm{opt}$) in each figure from (d) to (f) increases.

In Figs \ref{fig1Dpixel}(g-j), we compare how both bounds change as a function of source size $a$ for different temperature configurations. The black, red, and blue solid lines provide the QCRBs, and dashed black, dot-dashed red, and dotted blue provide the CRBs for single photon POVMs measurement for different $\mu$ (0.1, 0.5, 1.0), respectively. Further, the insets provide the total photon numbers in the detection modes. We observe that the blue dotted lines ($\mu=0.1$) are very close to the quantum limit and almost saturate the QCRBs for each source configuration for different  source sizes. Once we increase $\mu$, the gap between the two bounds increases as a function of source size $a$ due to the increased number of photons. For instance, compare the gap for black dashed lines ($\mu=1.0$) and blue dotted lines ($\mu=0.1$). This is due to the limitation of the single photon statistics for sources with total photon number greater than one ($\mathrm{Tr}(\Gamma)>1$).
\begin{figure}[t!]
	\centering
	\includegraphics[width=1.0\linewidth]{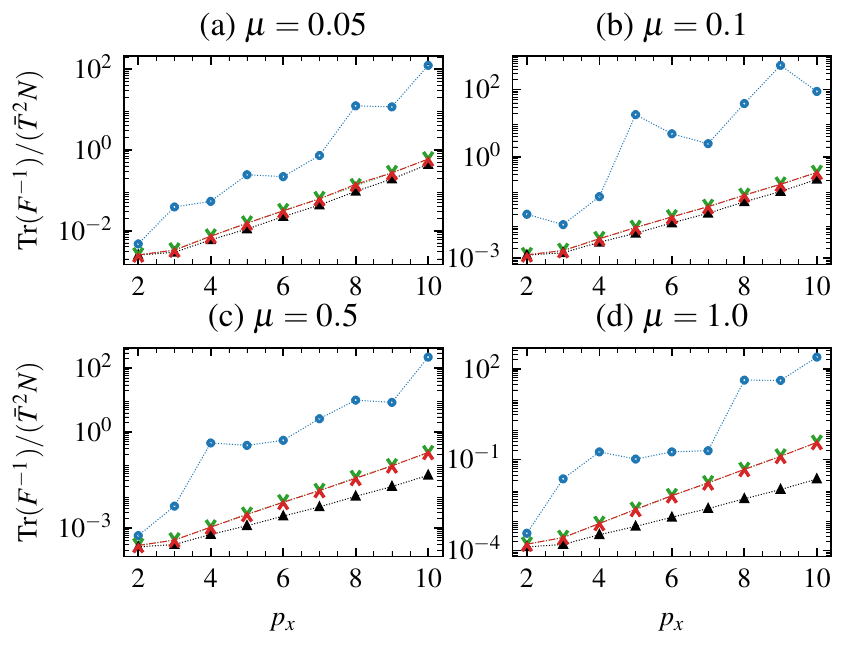}

	\caption{The scalar CRBs (dimensionless) for different numbers of pixels $p_x$ along the $\hat x$-axis 
          in a 1D array 
          and $\mu = (0.05,0.1,0.5,1.0)$ for figures (a) to (d), respectively. The black triangles represent QCRBs, and red upward wedges represent the scalar CCRBs that we get using the optimized unitary $\mathbf{U}_{\mathrm{opt}}^{\mathrm{image}}$ specific to the actual temperature distributions of source pixels. Green downward wedges are for the unitary $\mathbf{U}_{\mathrm{opt}}^{\mathrm{uniform}}$ optimized for uniform temperature 
          of the pixels used to estimate the actual temperature distribution
          with the same average temperature. 
          Blue circles correspond to scalar CRBs for the initial random unitary before optimization. Pixel size is $a=2.5$ km, average temperature  $\bar T = 300$ K, and sample size $N=10^6$.}
	\label{figpx}
\end{figure}

In general, the optimal unitary depends on the parameters (temperature distributions) we want to estimate. However, in real-life cases, we need to gain knowledge of the parameters to optimize the unitary completely. As we discuss in the section on two-pixel sources, a unitary for uniform temperature distributions can also be used to estimate different temperatures with the same $\bar T$ value. Experimentally, one can estimate the average temperature separately and construct the optimized unitary for the uniform temperature distribution ($\mathbf{U}_{\mathrm{opt}}^{\mathrm{uniform}}$). One then uses it to estimate the actual non-uniform temperature distribution. Further, we examine how both bounds change as a function of the number of pixels ($p_x$). In Fig.~\ref{figpx}, we show the CCRBs for different $\mu=(0.05,0.1,0.5,1.0)$ from (a) to (d), respectively. The blue circles represent the initial random unitary for the CG algorithm. The black triangles are the scalar QCRBs. The red upward wedges are the scalar CCRBs from the optimized unitary ($\mathbf{U}_{\mathrm{opt}}^{\mathrm{image}}$) specific to random temperature distributions of pixels. Further, the green downward wedges are for the optimized unitary for uniform temperature distributions ($\mathbf{U}_{\mathrm{opt}}^{\mathrm{uniform}}$) of the pixels, used to estimate the corresponding random unitary temperature distributions with the same pixel number and the same average temperatures. The bounds from $\mathbf{U}_{\mathrm{opt}}^{\mathrm{uniform}}$ (green wedges) and $\mathbf{U}_{\mathrm{opt}}^{\mathrm{image}}$ (red wedges) are very close to each other in this logarithmic scale. Also, both almost saturate the QCRBs for $\mu = 0.05$ and $\mu=0.1$ for different $p_x$. When we raise the number of pixels ($p_x$), we see that all bounds increase. Moreover, the gap between QCRBs and CCRBs from single photon measurements becomes more significant for $\mu=0.5$ and $\mu = 1.0$ compared to $\mu = 0.1$.

\subsection{Resolution of 2D sources}
\begin{figure}[t!]
	\centering
	\includegraphics[width=1.0\linewidth]{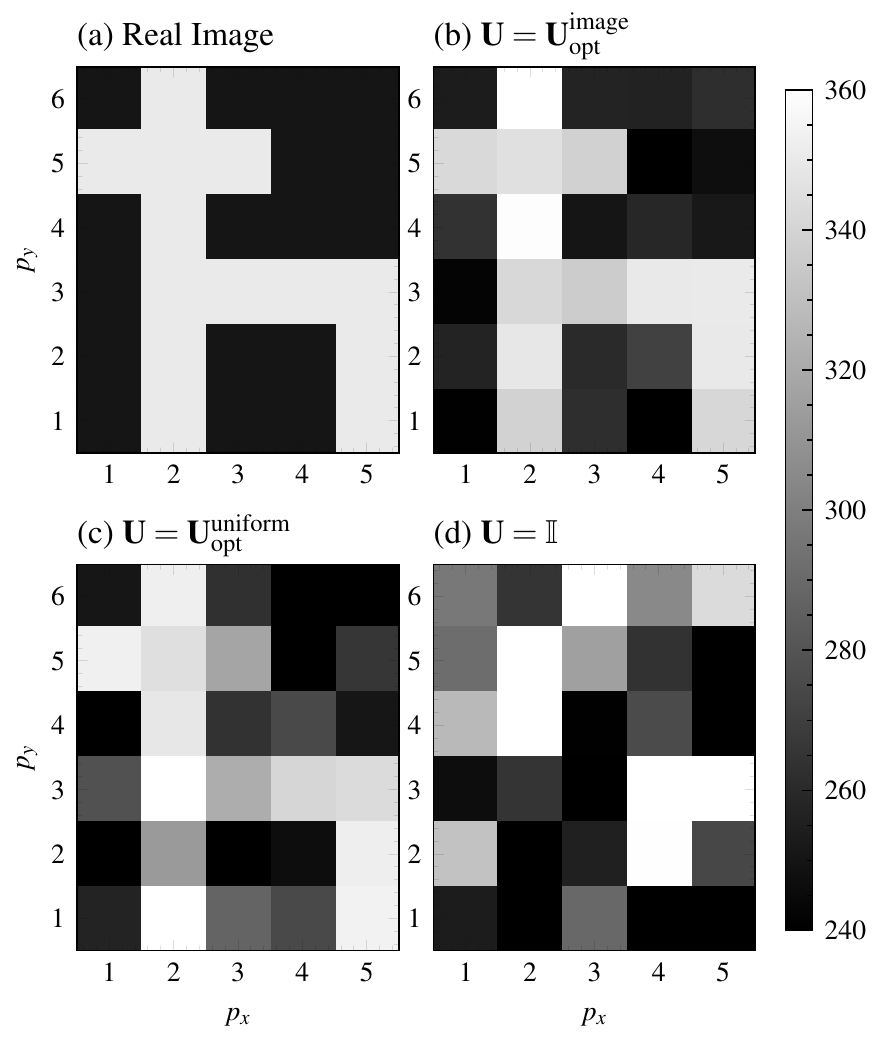}

	\caption{(a) The real image on the source plane with 30 pixels that will be estimated by using maximum likelihood estimator. (b) The reconstructed image after single photon detection in detection modes $\hat d_i $ obtained from using the optimized unitary $\mathbf{U}_{\mathrm{opt}}^{\mathrm{image}}$ specific to temperature distribution. (c) The reconstructed image using a unitary optimized for uniform temperature distribution $\mathbf{U}_{\mathrm{opt}}^{\mathrm{uniform}}$. (d) The reconstructed image using a local measurement of single photons considering $\mathbf{U} = \mathbb{I}$. Pixel size $a=3.0$ km, average temperature  $\bar T \sim 293$ K, and sample size $N=10^8$. }
	\label{figimage}
\end{figure}
This section considers an image with a total number of pixels $p = p_x p_y$ on the image plane. The number $n$ of the modes of the 2D array interferometers will be considered the same as $p$, with $n= n_x n_y$. The size of each pixel is set to $a = 3$ km, which is around
ten times smaller than the spatial resolution of SMOS considering van Cittert Zernike theorem, and the separation between the two nearest pixels is again set to zero. The parameters that we want to estimate are the temperatures of the 2D image $\bm{\theta} = \{ T_1, T_2,...T_{p}\}$. 
We consider the case of drastic photon losses 
and set $\mu=0.01$, which for $\bar T \sim 293$ K gives the total photon number around
$\mathrm{Tr}(\Gamma)\simeq 0.39$. 
In Fig.~\ref{figimage}(a), we consider an actual image of $\hbar$ using 30 pixels on the image plane and a 30 mode interferometer on the source plane. The unitary optimized ($\mathbf{U}_{\mathrm{opt}}^{\mathrm{image}}$) for this image or the unitary for a uniform temperatures distribution ($\mathbf{U}_{\mathrm{opt}}^{\mathrm{uniform}}$) is applied in the preprocessing stage to estimate the parameters. For the classical measurement, we consider a local measurement scenario with ($\mathbf{U} = \mathbb{I}$). Further, the image from different measurement strategies is reconstructed by using a maximum likelihood estimator for a sample of size $N$. In Fig.~\ref{figimage}(b), we reconstructed the image by using $\mathbf{U}_{\mathrm{opt}}^{\mathrm{image}}$. We have the advantage of the non-local measurement and the optimized unitary specific to the image. The reconstructed image is close to the actual image for this parameter regime. Though this unitary depends on the parameter set, we estimate that the same resolution limit may be achieved using an adaptive type of measurements \cite{fujiwara_strong_2011} by iteratively updating the unitary for each sample after measurement. 

However, this is beyond the scope  of this paper. On the other hand, for easy experimental realization, we reconstruct the image by using $\mathbf{U}_{\mathrm{opt}}^{\mathrm{uniform}}$ in Fig.~\ref{figimage}(c). One can independently estimate the average temperature from the source distribution and construct this general unitary for any image. As we see, the reconstructed image still reveals the actual image nicely, but as expected, it is not as sharp as the image from a specifically optimized unitary. We reconstructed the image from local measurement in Fig.~\ref{figimage}(d). Clearly, this reconstructed image is not close to the original one. This is expected for our pixel size $a= 3$ km, well below the limit of the Rayleigh resolution for SMOS, which is around 35 km, based on the van Cittert Zernike theorem \cite{anterrieu_resolving_2004,corbella_visibility_2004,levine_synthetic_1999,thompson_interferometry_2017}.

\section{Conclusion}\label{conclude}

In summary, we formulated passive remote sensing as a quantum
multi-parameter estimation problem, where we focused on the
temperatures on ground as parameters rather than geometrical
information of sources that are currently at the center of attention
in quantum imaging, such as the separation, centroid, or phases of
sources. An antenna array with as many antennas as desired pixels in
the source plane receives thermal electro-magnetic radiation in
receiver modes that are then mixed according to an optimized unitary
transformation. Single-photon detectors detect the photons in the
corresponding optimized detection modes. The function to be optimized
is a scalar classical Cram\'er-Rao bound, obtained by contracting the
inverse Fisher information matrix for estimating the temperatures from
the photon-counting results with a positive weight matrix.  With the
latter one can give different preferences for high resolution to
different parts of the image.  The optimization of the bound over all
unitary mode mixings leads to a ``most-informative bound''.  For a
uniform weight over all pixels we show that with this procedure one
can, in the case of the Gaussian white noise characteristic of thermal
states, approximatively saturate the scalar quantum Cram\'er-Rao bound
based on the contraction of the quantum Fisher information matrix for
the multi-parameter estimation problem with the same positive weight
matrix (chosen as the identity in the present work). In principle, the
optimized unitary depends on the actual temperature distribution, but
we showed that the unitary obtained from a uniform temperature
distribution gives still much better resolution than direct photon
counting in the incoming modes.   
For the optimization over the unitaries we used a conjugate gradient
algorithm. We showed that the found optimal mode mixing
followed by single photon detection leads to a spatial resolution of
the reconstructed images at least about an order of magnitude better
than Rayleigh's limit (about 3\,km instead of 35\,km for an antenna
array comparable with the one of SMOS, even for substantial photon
losses), given in the present case by the van Cittert-Zernike
theorem. The optimal unitary can be decomposed into $SU(2)$ group elements using beam splitters and phase shifters and can be realized
as linear optical quantum computing. Given the recent availability of
single-photon detection in the micro-wave domain, our results show a
path towards substantially enhanced resolution in passive remote
sensing compared to classical interferometers that essentially
implement homodyne quadrature measurements. Further improvements might
be possible for larger photon numbers or smaller losses if
photon-number resolved measurements are available.

\acknowledgements{DB and EK are grateful for support by the DFG,
  project number BR 5221\textbackslash3-1. We thank Gerardo Adesso for discussions, and DB thanks Yann Kerr, Bernard Roug\'e, and the
  entire SMOS team in Toulouse for
  valuable insights into that mission.}

\appendix

\section{Conjugate Gradient algorithm for optimization}
\label{appen}
This section summarizes a practical conjugate gradient (CG) algorithm given by Refs. \cite{abrudan_steepest_2008,abrudan_conjugate_2009,abrudan_efficient_2008}. The generic CG algorithm starts with ($k=0$) finding the conjugate gradient $\mathbf{G}_k$ of the cost function $F\left(\mathbf{U}_k\right) $ for an initial unitary matrix, where 
\begin{equation}
 \mathbf{G}_k=\frac{\partial}{\partial \mathbf{U}^*}F\left(\mathbf{U}_k\right).
\end{equation}
Then, the Riemannian gradient $\mathbf{W}_k$ at that point can be found by 
\begin{equation}
 \mathbf{W}_k=\mathbf{G}_k \mathbf{U}_k^\dagger-\mathbf{U}_k \mathbf{G}_k^\dagger .
\end{equation}
By determining the step size $\alpha$ using the Armijo method (see Ref. \cite{abrudan_efficient_2007} ) along the geodesic direction (in the direction of $-\mathbf{H}_{k}$), one can update the unitary by 
\begin{equation}
 \mathbf{U}_{k+1}= \exp(-\alpha \mathbf{H}_k)\mathbf{U}_{k} .
\end{equation}

Further, the new search direction can be found by using the Polak-Ribierre formula $\mathbf{H}_{k+1}=\mathbf{W}_{k+1} + \boldsymbol{\gamma}_k \mathbf{H}_{k}$, where 
\begin{equation}
 \boldsymbol{\gamma}_k :=\frac{\left\langle\mathbf{W}_{k+1}-\mathbf{W}_k , \mathbf{W}_k\right\rangle}{\left\langle\mathbf{W}_k, \mathbf{W}_k\right\rangle}.
\end{equation}
The inner product defined as $ \langle X,Y \rangle \equiv \mathrm{Tr}(X^\dagger Y)/2$ induces a bi-invariant metric on the unitary group $U(n)$. 
We reset the search direction periodically to ensure the direction of $\mathbf{H}_{k}$ is a descent direction. Then the next iteration continues accordingly (see pseudo-code in Algorithm \ref{algorithm}). The algorithm runs until it converges to a minimum value of the cost function or a maximum number of iterations $k_{\mathrm{max}}$. To efficiently deal with the gradient of the cost functions, we used the PyTorch gradient function. PyTorch is used in machine learning for it is GPU capabilities.

\begin{figure}[h!]
	\begin{algorithm}[H]
		\caption{Conjugate gradient algorithm for unitary optimization}
	\begin{algorithmic}[1]
		\Require $k=0$, $\mathbf{U}_k =$ Random Unitary, $n= \mathbf{dim}(\mathbf{U}_k)$, $\alpha =1$
		\While{$k\neq k_{\mathrm{max}}$}
			\If{$k$ modulo $n^2=0$}
				\State $\mathbf{G}_k=\frac{\partial}{\partial \mathbf{U}^*}F\left(\mathbf{U}_k\right) $
				\State $\mathbf{W}_k=\mathbf{G}_k \mathbf{U}_k^\dagger-\mathbf{U}_k \mathbf{G}_k^\dagger $
				\State $\mathbf{H}_k:=\mathbf{W}_k$
			\Else 
				\State $\mathbf{W}_{k} \gets \mathbf{W}_{k+1}$
				\State $\mathbf{H}_{k} \gets \mathbf{H}_{k+1}$
			\EndIf
			\\
			\State $\mathbf{P}_{k} = \exp(-\alpha \mathbf{H}_k)$
			\State $\mathbf{Q}_{k} = \mathbf{P}_{k}\mathbf{P}_{k}$
			\\
			\While{$F(\mathbf{U}_k)-F(\mathbf{Q}_k\mathbf{U}_k) \geq  \alpha  \left\langle\mathbf{W}_k, \mathbf{H}_k\right\rangle $ }
				\State $\mathbf{P}_{k} = \mathbf{Q}_{k}$
				\State $\mathbf{Q}_{k} = \mathbf{P}_{k}\mathbf{P}_{k}$
				\State $\alpha =2\mu$
			\EndWhile
			\\
			\While{$F(\mathbf{U}_k)-F(\mathbf{P}_k\mathbf{U}_k) <  (\alpha/2)  \left\langle\mathbf{W}_k, \mathbf{H}_k\right\rangle $ }
				\State $\alpha =\alpha/2$
				\State $\mathbf{P}_{k} = \exp(-\alpha \mathbf{H}_k)$
			\EndWhile
			\\
			\State $\mathbf{U}_{k+1}=\mathbf{P}_{k}\mathbf{U}_{k} $
			\State $\mathbf{G}_{k+1}=\frac{\partial}{\partial \mathbf{U}^*}F\left(\mathbf{U}_{k+1}\right) $
			\State $\mathbf{W}_{k+1}=\mathbf{G}_{k+1} \mathbf{U}_{k+1}^\dagger-\mathbf{U}_{k+1} \mathbf{G}_{k+1}^\dagger $
			\State $\boldsymbol{\gamma}_k :=\frac{\left\langle\mathbf{W}_{k+1}-\mathbf{W}_k , \mathbf{W}_k\right\rangle}{\left\langle\mathbf{W}_k, \mathbf{W}_k\right\rangle}$
			\State $\mathbf{H}_{k+1}=\mathbf{W}_{k+1} + \boldsymbol{\gamma}_k \mathbf{H}_{k}$
			\\
			\If{$\left\langle\mathbf{W}_{k+1}, \mathbf{H}_{k+1}\right\rangle \leq 0 $ }
			\State $\mathbf{H}_{k+1}=\mathbf{W}_{k+1}$
			\EndIf
			\\
	
			\State $\mathbf{U}_{k} \gets \mathbf{U}_{k+1} $
			\State $k \gets k+1$
		\EndWhile 
	\end{algorithmic}
	\label{algorithm}
	\end{algorithm}
	\end{figure}

\newpage
\vfill
\bibliography{bibliography}

\end{document}